\documentclass[conference]{IEEEtran}

\usepackage{booktabs}
\usepackage{amsmath,amssymb}
\usepackage{todonotes}
\usepackage{url}

\usepackage{breakurl}
\usepackage{amsmath}
\usepackage{algpseudocode}
\usepackage{algorithm}
\usepackage{hhline}
\usepackage{graphicx}
\graphicspath{ {figures/} }
\usepackage{svg}
\usepackage{subfig}
\usepackage{float}

\begin{document}
%
\title{\textbf{BeatCoin: Leaking Private Keys from Air-Gapped Cryptocurrency Wallets}}

\author{\IEEEauthorblockN{Dr. Mordechai Guri}
\IEEEauthorblockA{Ben-Gurion University of the Negev, Israel\\Cyber-Security Research Center\\
gurim@post.bgu.ac.il\\demo video (1): \url{https://youtu.be/2WtiHZNeveY}\\demo video (2): \url{https://youtu.be/ddmHOvT866o}}}


%


\maketitle

\begin{abstract}
Cryptocurrency wallets store the wallet’s private key(s), and hence, are a lucrative target for attackers. With possession of the private key, an attacker virtually owns all of the currency in the compromised wallet. Managing cryptocurrency wallets offline, in isolated ('air-gapped') computers, has been suggested in order to secure the private keys from theft. Such air-gapped wallets are often referred to as 'cold wallets.'

In this paper we show how private keys can be exfiltrated from air-gapped wallets. In the adversarial attack model, the attacker \textit{infiltrates} the offline wallet, infecting it with malicious code. The malware can be preinstalled or pushed in during the initial installation of the wallet, or it can infect the system when removable media (e.g., USB flash drive) is inserted into the wallet's computer in order to sign a transaction. These attack vectors have repeatedly been proven feasible in the last decade  (e.g., \cite{25ofneww53:online},\cite{NewCrypt8:online},\cite{W32Dapro94:online},\cite{Sporathe68:online},\cite{CiscosTa82:online},\cite{ShadowPa49:online},\cite{Bewareof42:online},\cite{Kaspersk99:online},\cite{AFannyEq7:online},\cite{langner2011stuxnet}). Having obtained a foothold in the wallet, an attacker can utilize various air-gap covert channel techniques  (\textit{bridgeware} \cite{Guri:2018:BAM:3200906.3177230}) to jump the air-gap and exfiltrate the wallet’s private keys. We evaluate various exfiltration techniques, including physical, electromagnetic, electric, magnetic, acoustic, optical, and thermal techniques. This research shows that although cold wallets provide a high degree of isolation, it’s not beyond the capability of motivated attackers to compromise such wallets and steal private keys from them. We demonstrate how a 256-bit private key (e.g., bitcoin's private keys) can be exfiltrated from an offline, air-gapped wallet of a fictional character named Satoshi within a matter of seconds\footnote{demonstration video: https://cyber.bgu.ac.il/advanced-cyber/airgap}.    
\end{abstract}


%
\IEEEpeerreviewmaketitle

\section{Introduction}
Cryptocurrencies such as bitcoin \cite{BitcoinO22:online} and Etherum \cite{Ethereum92:online} have emerged as a popular medium of money exchange, with a large associated ecosystem and supporting community. In a nutshell, cryptocurrencies can be
considered as a decentralized payment network
that is maintained by its users without the need for a single authority. A global log known as 'blockchain' records all of the transactions in the network. Each block in the blockchain represents a number of transactions and includes the transaction data, a timestamp, and a cryptographic hash of the previous block. The distributed nature of the blockchain makes it resistant to adversarial tampering of the information contained in its logs, offering level of protection that is inherently not possible with standard centrally managed databases. The blockchain technology has also been adopted by many other applications such as smart contracts \cite{kosba2016hawk}, medical records \cite{azaria2016medrec} and digital voting \cite{pilkington201611}.

As of the time of this writing (April 2018), more than 3000 different cryptocurrencies are available on the Internet. Most cryptocurrencies share the technology and implementation of the larger cryptocurrencies like bitcoin.
 
The scope of this paper is relevant to all cryptocurrencies and blockchain applications (e.g., smart contracts), although in this paper we will largely focus on bitcoin, which is the most popular cryptocurrency today.

\subsection{Private \& Public Keys}
The whitepaper describing bitcoin was published in 2008 by an unknown person (or people) named 'Satoshi Nakamoto.' The paper ("Bitcoin: A Peer-to-Peer Electronic Cash System" \cite{nakamoto2008bitcoin}) was published on a cryptography mailing list and described the bitcoin network principles. In bitcoin architecture, the payments are performed by issuing \textit{transactions} describing the currency transfers between two \textit{peers} in the network. Every peer in the bitcoin network is referred to by a unique number called a \textit{bitcoin address}. Each bitcoin address is associated with a public key and a private key. The public key is a 65 byte number and the private key is a 32 byte number (256-bit). The public keys are published in the bitcoin network and they are publicly available. Transactions, which are \textit{signed} by a private key can be \textit{verified} by anyone using the corresponding public key. The detailed process of performing transactions in the bitcoin network is provided in the original whitepaper \cite{nakamoto2008bitcoin}.

Although there are various cryptocurrencies with different cryptographic schemes and key sizes, the most popular cryptocurrencies use 256-bit private keys. Table \ref{tabletop5} lists the top-10 cryptocurrencies\footnote{By market capitalization, according to https://coinmarketcap.com/ (April 2018)} and the size of their private keys. 

\begin{table}[]
	\centering
	\caption{The top-10 cryptocurrencies/platforms and the size of their private keys}
	\label{tabletop5}
	\begin{tabular}{@{}lll@{}}
		\toprule
		Symbol & Cryptocurrency & Private key \\ \midrule
		BTC    & Bitcoin   & 256-bit     \\
		ETH    & Ethereum  & 256-bit     \\
		XRP    & Ripple         & 256-bit     \\
		BCH    & Bitcoin cash   & 256-bit     \\
		LTC    & Litecoin       & 256-bit     \\
		EOS    & EOS            & 256-bit     \\
		ADA    & Cardano        & 256-bit     \\
		XLM    & Stellar        & 256-bit     \\
		NEO    & NEO            & 256-bit     \\
		MIOTA  & IOTA           & 256-bit     \\ \bottomrule
	\end{tabular}
\end{table}

\subsection{Cryptocurrency Wallets}
A cryptocurrency wallet is a virtual object which refers to the digital credentials of the currency holdings, and it is essentially the public and private keys associated with a peer. A bitcoin wallet contains one or more private keys, which are mathematically related to the bitcoin addresses generated for the wallet. Private keys are the most valuable asset in a wallet as they can be used to transfer all bitcoins in a wallet to another peer. They must be be kept secured and safe to avoid theft and lost. 

Some bitcoin wallet applications use a single seed to generate many pairs of public and private keys. This approach is called a hierarchical deterministic (HD) wallet. In one of the common implementations of this type of wallet, the seed value consists of a random 128-bit value represented as a 12 word mnemonic using common English words.

\subsection{Types of Wallets}
There are different approaches for managing cryptocurrency wallets. At a technical level, they can be categorized into software wallets, hardware wallets, and paper/brain wallets.

\subsubsection{Software wallets}
A software wallet is the application which stores the public and private keys. It also manages
the bitcoin transactions, allowing clients to send bitcoins and view their balance. Most of the software wallets today provide a user-friendly control panel to view the wallet's status and perform online transactions. There are several types of software wallets, and the most important of them are listed below. 

\begin{itemize}
	\item {{Client-side wallets.}} Client-side wallets are applications that the user installs on his/her PC, tablet, or smartphone. The public and private keys are stored locally in a wallet file. Many client-side wallet applications support maintaining different types of cryptocurrencies.  
	\item {{Web-based wallets.}} Web-based wallets are managed by trusted third parties and can be accessed via online websites. The private keys are stored in the provider's database and are not exposed to the client side. 
	\item {{Watch-only wallets.}} Watch-only wallets allow the user to track existing transactions but don't allow them to initiate new ones. Only the public keys are stored in the wallets.
    \item {{Cold ('air-gapped') wallets.}} Cold wallets are managed offline, disconnected from the Internet. Unlike online wallets (hot wallets), cold wallets are not connected to the bitcoin network and hence, can not initiate online transactions. Since cold wallets are managed offline, usually on an air-gapped computer, the private keys are protected from online threats and thought to be safe from cyber theft. Air-gapped wallets will be discussed in Section \ref{sec:2} and Section \ref{sec:3}.   
\end{itemize}

\subsubsection{Hardware wallets}
In hardware wallets the private keys are stored in dedicated trusted hardware modules. They are connected to the host computer via USB interface and commonly contain security features such as PIN codes and embedded screens. In hardware wallets the transactions are signed within a trusted computational environment in the hardware (e.g., the ARM TrustZone), and the private keys are not exposed to the host computer. The signed transactions are delivered to the wallet application via a specific API provided by the vendor of the hardware wallet. Hardware wallets are less vulnerable to online attacks because the private keys can not be accessed by malware in the host computer.Known hardware wallets include TREZOR \cite{TREZORBi21:online} and Ledger Nano S \cite{LedgerWa93:online}. 
    
\subsubsection{Paper and brain wallets}
In a paper wallet the private keys are kept on a printed piece of paper. They are commonly printed  in a form of alphabet string or encoded as a QR code. There are online websites that generate printable wallets (e.g., www.bitaddress.org). Paper wallets are considered the most secure, because they are completely offline and are thus, largely unexposed to cyber threats. Similar to paper wallets, in  a brain wallet the private keys are not stored in digital form. Instead, the wallet owner memorizes the wallet’s mnemonic recovery phrase. If the mnemonic recovery phrase are forgotten, the bitcoins are lost.

\subsection{Wallet Security}
The security of a wallet is correlated directly with the level of security of its private keys. Hot wallets are always online and hence, vulnerable to cyber-attacks. Attackers can inject a malicious code into the host computer running the wallet application using wide range of techniques including: compromised web-sites \cite{provos2007ghost}, drive-by-download \cite{cova2010detection}, malvertising \cite{sood2011malvertising}, social engineering \cite{peltier2006social}, malicious documents \cite{smutz2012malicious}, and so on. A malware in the host can easily access the file that stores the private key(s) and leak them to a remote attacker via the Internet. Several cryptocurrency stealing malware have been found in the wild recently: ComboJack \cite{SureI’ll82:online}, CryptoShuffler \cite{CryptoSh40:online}, and TrickBot \cite{TrickBot58:online}. Such online attacks are unavoidable as long as the wallet is connected to the Internet.

Hardware wallets are physically connected to online computers (e.g., when transactions are initiated) and can be considered hot wallets. However, the trusted hardware and secure design provide  \textit{logical isolation} of the private keys, preventing malicious code from accessing them. Note that hardware wallets don't provide hermetic security. In recent years bugs and vulnerabilities were found in the implementation of hardware components \cite{kocher2018spectre,lipp2018meltdown}, including in trusted execution environments like the ARM TrustZone \cite{us15Shen95:online} \cite{ProjectZ70:online}. These types of vulnerabilities allow attackers to evade hardware-enforced isolation mechanisms and access protected data.  

Air-gapped wallets are thought to provide the highest level of protection of the private keys - since the private keys are kept in an offline computer, they are \textit{physically} isolated from the Internet and hence, cannot be accessed by hackers and leaked out.  

Table \ref{attacks3} presents the four types of wallets along with the level of isolation they provide and the attack surface for each wallet. 


\begin{table*}[t]
	\centering
	\caption{The level of isolation and attack surface of the private keys}
	\label{attacks3}
	\begin{tabular}{@{}lll@{}}
		\toprule
		Wallet type             & Isolation                             & Attack surface                                                                                     \\ \midrule
		Hot wallets              & No isolation                          & Online attacks (e.g., ComboJack \cite{SureI’ll82:online}, CryptoShuffler \cite{CryptoSh40:online} and TrickBot \cite{TrickBot58:online})                                                                                    \\
		Hardware wallets        & Logical isoaltion (hardware enforced) & \begin{tabular}[c]{@{}l@{}}Hardware implementation bugs and vulnerabilities (e.g.,\cite{us15Shen95:online} \cite{ProjectZ70:online})\end{tabular} \\
		Air-gapped cold wallets    & Physical isolation                    & \begin{tabular}[c]{@{}l@{}}Air-gap infiltration and exfiltration  (this paper)\end{tabular}          \\
		Paper wallet, brain wallet & Physical isolation                    & Physical lost, theft, forgetting they mnemonic phrase, death, etc.                                                                  \\ \bottomrule
	\end{tabular}
\end{table*}

In this paper we focus on the vulnerability of air-gapped wallets to cyber-attacks. In particular, we show that despite the level of isolation, private keys can be exfiltrated from such wallets to the Internet. First, we discuss the methods that can be used by attackers to infiltrate the air-gapped wallets. Second, we show that attackers can exfiltrate private keys over the air-gap using special types of covert channels.

\section{Wallet Infiltration}
\label{sec:2}
In this section we present techniques which can be used by attackers to compromise air-gapped wallets and infect them with malware. We also show that the infiltration of a wallet can be done at a very early stage, before the wallet software installed in the system and before the private keys are generated.

\subsection{Post-Installation}
Although air-gapped wallets are kept offline, there are occasions when external media is inserted into the air-gapped host. This media might be a USB flash drive, an optical disk (CD/DVD), or a memory card (SD card). 
The most common scenario of introducing removable media to air-gapped wallets involves signing and broadcasting transactions.
Signing transactions and distributing them online is commonly done through an external USB flash drive. For example, in the Electrum bitcoin client, signing a transaction in a cold wallet is done via a file saved in a removable media device \cite{ColdStor29:online}. Once the transaction is signed offline, the transaction file is moved to the online wallet and broadcasted over the bitcoin network. The same work flow is true for other wallet applications as well \cite{BitKeySe86:online}. 

The removable media transfers between online and offline wallets can be used by attackers to infiltrate air-gapped wallets and infect them with malware. Using removable media (especially USB flash drives) to spread malware across PCs is known to be effective. Research on this topic released by PandaLabs \cite{25ofneww53:online} stated that 25\% of all worms in 2010 relied on USB devices to spread to other PCs. Out of 10,000 firms infected with malware, more than 2,500 reported that the attack had originated with an infected USB flash drive.  Malware such as Daprosy \cite{W32Dapro94:online}, CryptoLocker \cite{NewCrypt8:online}, Spora, \cite{Sporathe68:online} and ZCrypt \cite{Bitdefen74:online} used removable drives as a primary spread vector. In the arena of advanced threats, many famous APTs used removable media to infiltrate air-gapped systems, including ProjectSauron \cite{TheProje91:online}, Fanny \cite{AFannyEq7:online}, Regin \cite{Kaspersk99:online}, Stuxnet \cite {langner2011stuxnet} and Agent.Btz \cite{grant2009cyber}. HammerDrill 2.0, disclosed in WikiLeaks in 2017, is a cross-platform attacking tool that can use CD/DVD as a covert-channel to compromise air-gapped systems \cite{HammerDr64:online}. The Brutal Kangaroo framework, \cite{WikiLeak41:online} also disclosed in WikiLeaks in the same year, includes components which enable the infection of closed networks via USB devices. In April 2018, researchers exposed a file system vulnerability (CVE-2018-6791) which allows arbitrary command execution on Linux systems via external thumb drives \cite{httpswww55:online}.

\begin{figure}[t]
	\centering
	\includegraphics[width=\linewidth]{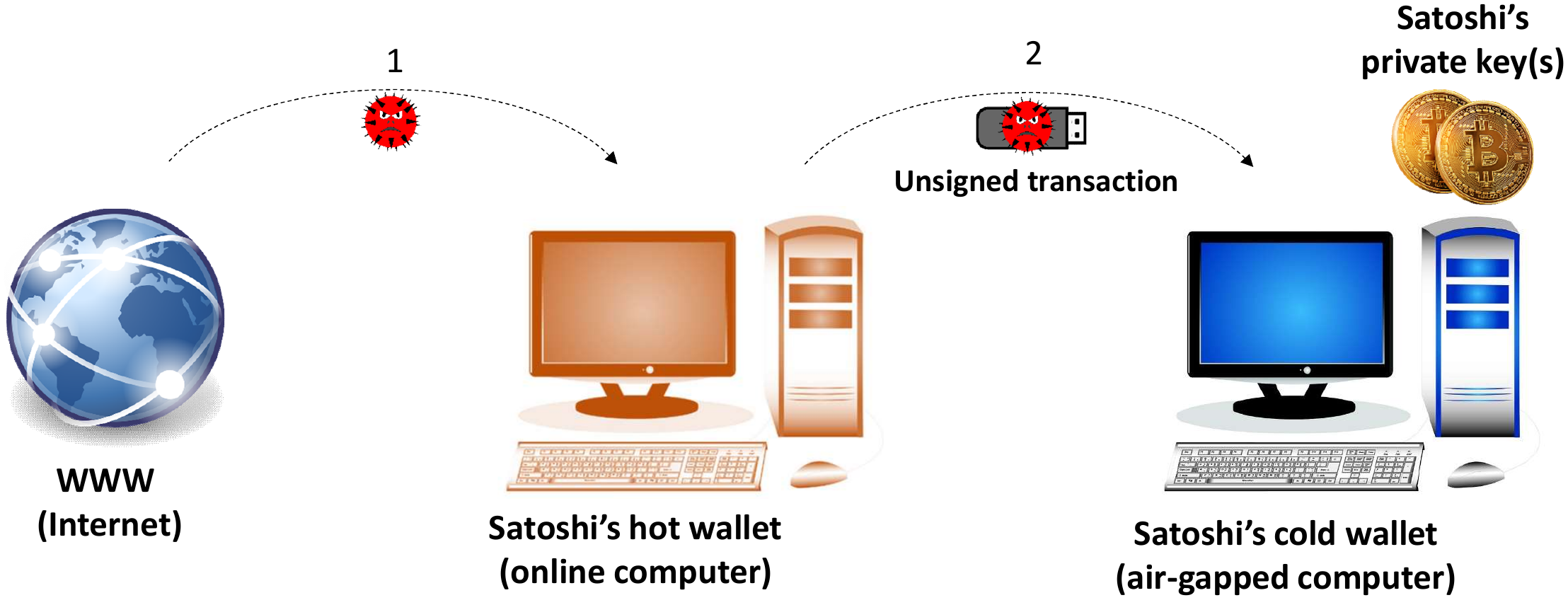}
	\caption{\textbf{Infiltration of an air-gapped wallet during the transaction signing process. When Satoshi plugs the USB flash drive into the air-gapped computer the system is infected.}}
	\label{fig:1}
\end{figure}

The use of such vulnerabilities and tools enables hackers to compromise air-gapped wallets. The infiltration process is illustrated in Figure \ref{fig:1}. In the initial stage the Internet-connected computer of the wallet owner is infected with a malware. Once removable media is inserted into the online computer (e.g., to copy the unsigned transaction file), it becomes infected with malware. When the removable media is inserted into the air-gapped computer, it then infects the air-gapped system.

\subsection{Pre-Installation}
The air-gapped computer might be compromised even before the wallet is installed, via an infected operating system (OS) or compromised image of the wallet software. 

\subsubsection{Modified OS distribution / modified wallet}
Attackers can modify OSs and wallets on the download sites. In a famous attack that occurred in 2016, hackers modified the Linux Mint image file (ISO), inserted a backdoor into it, and managed to hack the official website to point to the compromised image \cite{Bewareof42:online}. In the same way, instances of wallet software might be distributed with a built-in malware. Such attacks were shown to be feasible in 2017, when an official version of CCleaner was compromised and distributed with a built-in backdoor \cite{CCleaner9:online}.
\subsubsection{Post-download infection}
A cold wallet is commonly installed in the air-gapped computer using an OS and a wallet application that were downloaded from the Internet. They are then uploaded to removable media (e.g., USB flash drive) and installed on the air-gapped computer. Malware can infect the removable media or the wallet image after the downloads and just before its installation in the air-gapped computer.

Table \ref{infvec} lists the attack vectors for air-gapped wallets. Note that there are additional attack vectors such as supply chain attacks and physical access \cite{mcfadden2010supply} which can be used for infiltration. However, because such attacks are often targeted, and require a significant amount of funding and resources, they require, we consider them less relevant threats for private cryptocurrency wallets. 

\begin{table}[]
	\centering
	\caption{Infiltration vectors}
	\label{infvec}
	\begin{tabular}{@{}lll@{}}
		\toprule
		Infiltration vector         & Infection method/stage                                                                                                 & Examples of past attacks \\ \midrule
		Removable media             & \begin{tabular}[c]{@{}l@{}}Wallet installation/\\ Transactions signing\end{tabular}   & 
		\cite{W32Dapro94:online},\cite{NewCrypt8:online},\cite{Sporathe68:online},\cite{Bitdefen74:online},\cite{TheProje91:online},\cite{AFannyEq7:online},\cite{Kaspersk99:online}   \\
		Modified images & \begin{tabular}[c]{@{}l@{}}Modified ISO/\\ Compromised websites/\\ Post-download infection\end{tabular} & \cite{Bewareof42:online},\cite{CCleaner9:online},\cite{ShadowPa49:online},\cite{CiscosTa82:online}                    \\ \bottomrule
	\end{tabular}
\end{table}

\section{Keys Exfiltration}
\label{sec:3}
Having a foothold in the air-gapped computer running the wallet, allows an attacker to utilizes air-gap covert channels to leak the private keys out.
Air-gap covert channels are special covert channels that enable communication with air-gapped computers - mainly for the purpose of data exfiltration. In 2018, Guri coinded the term \textit{bridgeware} \cite{Guri:2018:BAM:3200906.3177230} to refer to the class of malware that exploits air-gap covert channels in order to bridge the air-gap between isolated computers/networks and attackers. The air-gap covert channels can be classified into seven main categories which are discussed in the context of the current attack model (air-gapped wallets) in this section: physical, electromagnetic, electric, magnetic, acoustic, optical, and thermal.

In this type of attack vector the wallet keys are transmitted from the offline wallet to a nearby (online) computer, smartphone, webcam, or other type of receiver via these covert channels. The private keys are then sent to the attacker through the Internet. In the next subsections, we discuss these covert channels and examine the security threat they pose to cryptocurrency wallets.

\subsection{Physical (Removable Media)}
As discussed in Section \ref{sec:2}, although cold wallets are physically disconnected from the Internet, a removable media device (e.g., USB flash drive or CD/DVD) may be inserted into the air-gapped host. Attackers can use this as an opportunity for exfiltrating private keys. Note although such occasions might be rare, one is enough for an attacker to leak one or several private keys. 

The most common scenario for the use of removable media is for offline transaction signing. After a transaction is signed in the offline computer, it must be broadcasted to the bitcoin network. This can be done by transferring the signed transaction file to the online wallet computer through a USB flash drive \cite{ColdStor29:online}.
For example, the \textit{bitcore-wallet} manual for air-gapped wallets states that \textit{"Transactions can be pulled from BWS using a proxy device, then downloaded to a pendrive to be moved to the air-gapped device, signed there, and then moved back the proxy device to be sent back to BWS. Note that Private keys are generated off-line in the airgapped device."} \cite{GitHubbi58:online}.
 
Using removable media to maintain covert channels is a known techniques used by malware and worms \cite{25ofneww53:online}. The HammerDrill \cite{HammerDr64:online} and Brutal Kangaroo \cite{Wikileak92:online} frameworks disclosed in WikiLeaks in 2017 are capable of exchanging data with closed networks via removable media. Similarly, the ProjectSauron APT \cite{TheProje91:online} is capable of exfiltrating data from air-gapped networks via USB sticks. The same mechanism exists in Equation \cite{Equation51:online}, Regin \cite{Kaspersk99:online} and Fanny APTs \cite{AFannyEq68:online}. In the case of Fanny, the APT creates a hidden storage area in the USB flash drive, collects the system information, and saves it in the hidden area. When the USB flash drive was inserted into an Internet-connected computer the data was exfiltrated.

\begin{figure}[t]
	\centering
	\includegraphics[width=\linewidth]{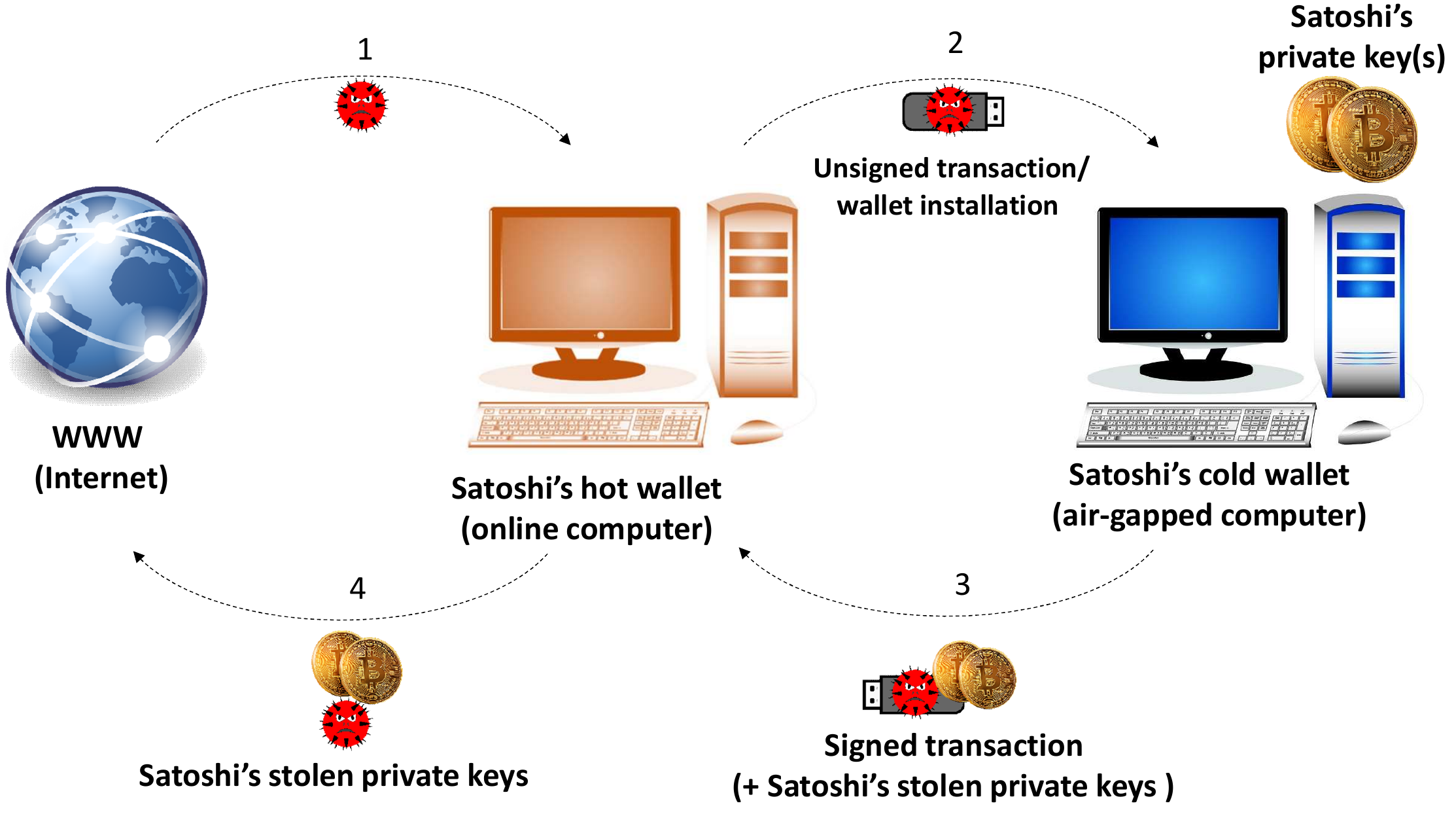}
	\caption{\textbf{\textbf{Exfiltration of the private keys during the transaction signing process. When Satoshi plugs the USB flash drive into the air-gapped wallet, the private keys are stolen.}}}
	\label{fig:2}
\end{figure}

This attack vector is illustrated in Figure \ref{fig:2}. When a USB flash drive is inserted into the air-gapped computer (e.g., for signing a transaction), the malware stores the private key(s) in a hidden file/partition. Once the USB flash drive is inserted into the hot wallet computer (e.g., for broadcasting the signed transaction), the malware read the private keys and sends it to the attacker over the Internet. Note that the extra I/O operations of writing the private keys to the file-system in the flash drive have a negligible effect in terms of time and are virtually unnoticeable by the user.

\subsection{Electromagnetic}
Electromagnetic based covert channels have been studied since the 1990s. Back in 1998, Kuhn et al showed that it is possible to generate electromagnetic emissions from a PC's display cables \cite{kuhn1998soft}. They also showed that binary information can be modulated on top of the emitted signals. Based on this work, Thiele \cite{Tempestf48:online} presented a program which uses the computer monitor to transmit AM radio signals modulated with audio. He demonstrated the method by transmitting the tune Beethoven piece, "Letter to Elise," and showed how it could be heard from a simple radio receiver located nearby. Although the existence of electromagnetic covert channels has long been known, since a radio receiver needs to be located close to the emanating computer, this covert channel was considered less practical for cyber-attacks.

\subsubsection{AirHopper}
More recently, Guri et al demonstrated AirHopper \cite{guri2014airhopper,guri2017bridging}, a malware that is capable of exfiltrating data from air-gapped computers to a nearby smartphone via FM signals emitted from the screen cable. The covert transmissions are received by the FM radio receiver which is integrated into many modern smartphones. They also discussed stealth and evasion techniques that help hide the malicious transmission. In a case of an AirHopper attack, the effective distance is a few meters from the air-gapped computer, and the effective bit rate is 100-480 bit/sec. The AirHopper attack can be used to leak the private keys from an air-gapped wallet to the user's smartphone in a few seconds.

\begin{figure}[t]
	\centering
	\includegraphics[width=\linewidth]{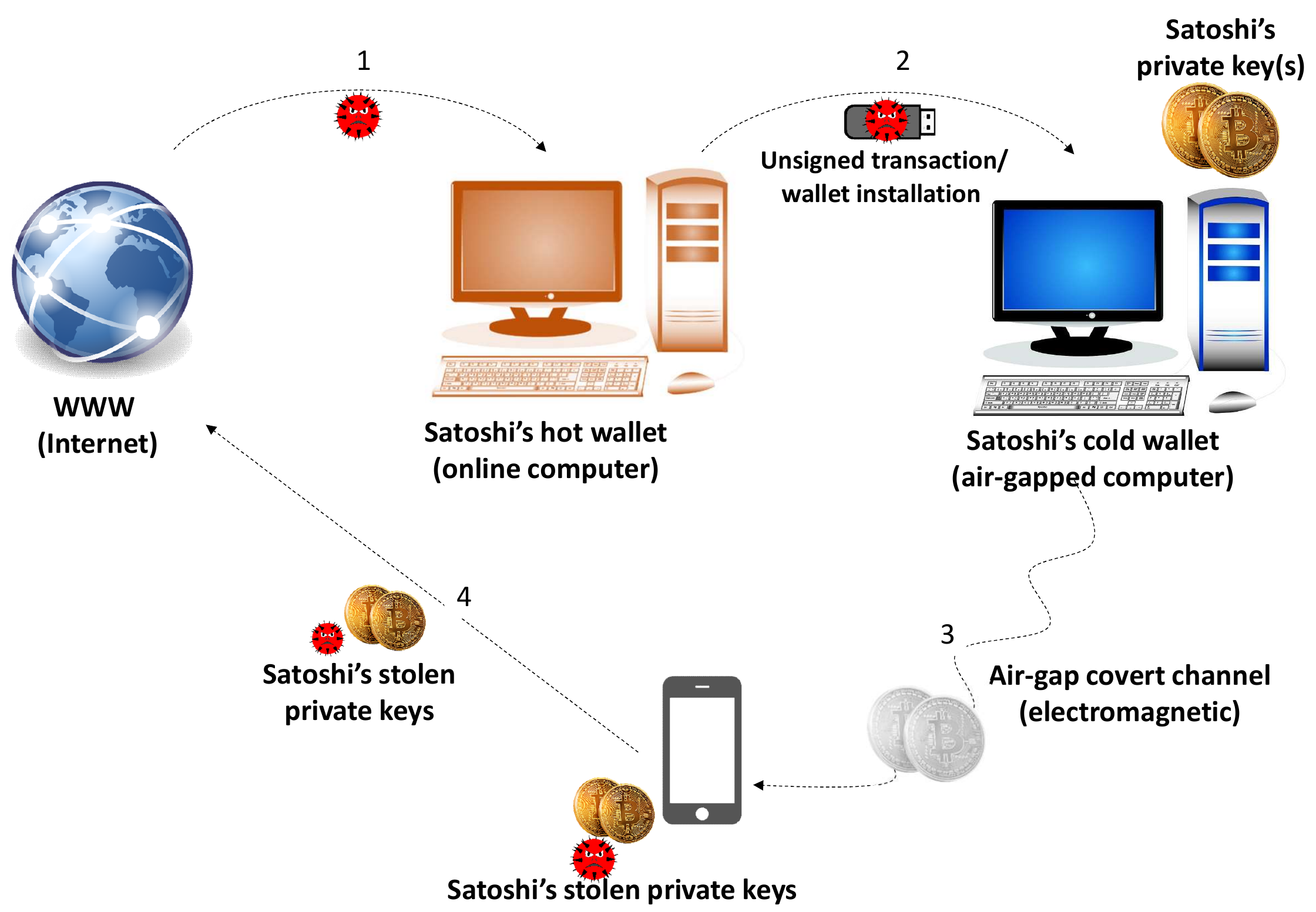}
	\caption{\textbf{Exfiltration of the private keys via electromagnetic covert channels. Satoshi's private keys are transmitted to the nearby smartphone via electromagnetic signals (e.g., AirHopper \cite{guri2014airhopper}, GSMem \cite{guri2015gsmem}, RADIoT \cite{RADIoTguri}) and sent to the attacker through the Internet.}}
	\label{fig:3}
\end{figure}
 
\subsubsection{GSMem}
Similar to AirHopper, the GSMem attack \cite{guri2015gsmem}, enables leaking the data from air-gapped wallets to nearby mobile phones. In this technique malware generates interferences in the cellular bands of the GSM, UMTS, and LTE specification. The signals are generated from the buses which connect the RAM and the CPU on the motherboard. The transmission can be received by a rootkit hidden in the baseband firmware of a nearby mobile phone. In a case of a GSMem attack, the mobile phone must be located close to the air-gapped computer, and the effective bandwidth is 1-2 bit/sec. The GSMem attack can be used to leak the private keys from an air-gapped wallet to the user's smartphone in a few minutes. 

%

\subsubsection{RADIoT}
In the RadIoT attack \cite{RADIoTguri} data can be leaked from air-gapped embedded systems and IoT devices via radio signals. The radio signals - generated from various buses and general-purpose input/output (GPIO) pins of the embedded devices - can be modulated with binary data. In this case, the transmissions can be received by an AM or FM receiver located nearby the device. This attack is relevant to cases where the air-gapped wallet is maintained in embedded and low-power devices, such as a Raspberry PI as suggested in \cite{SecureYo55:online}\cite{Offlineb75:online}.In the case of a RADIoT attack, the private keys can be exfiltrated at bit rate of tens to hundreds of bits per second, depending on the type of device used. The RADIoT attack can be used to leak the private keys from an air-gapped wallet to the user's smartphone or RF receiver in a few seconds. 

The electromagnetic based covert channels are illustrated in Figure \ref{fig:3}. In this case, Satoshi's private keys are transmitted to the nearby smartphone via electromagnetic signals (e.g., AirHopper \cite{guri2014airhopper}, GSMem \cite{guri2015gsmem}, RADIoT \cite{RADIoTguri}), and sent to the attacker through the Internet.

\subsection{Electric}
In 2018, Guri et al presented PowerHammer \cite{2018powerhammer}, an attack which can be used to exfiltrate data from air-gapped computers through power lines. A malware in the air-gapped computer controls the power consumption of the system by changing the CPU workload. It encodes data on top of the changes in current flow, which is propagated through the power lines. In this work, the authors presented a type of attack named phase level power-hammering in which the attacker probes the power lines at the phase level in the main electrical service panel. In the phase level attack, they were able to exfiltrate data at a bit rate of 10 bit/sec. This attack requires the attacker to obtain physical access to the electrical service panel where the air-gapped computer is located. The PowerHammer attack can be used to leak the private keys from an air-gapped wallet in just a few seconds or minutes. 
 
\subsection{Magentic}
The private keys can be leaked from air-gapped wallets via magnetic fields.

\subsubsection{ODINI and MAGNETO}
The ODINI \cite{guri2018odini} and MAGNETO \cite{guri2018magneto} attacks enable the exfiltration of data via magnetic signals generated by the computer processors. Magnetic signals can also be generated from the reading/writing heads of hard disk drives \cite{matyunin2016covert}. The receiver may be a magnetic sensor or a smartphone located near the computer. One of the interesting properties of ODINI and MAGNETO attacks is that the low frequency magnetic fields can bypass Faraday shielding. Thus, in the case on an air-gapped wallet, private keys can be exfiltrated even if the wallet or receiver smartphone is enclosed within a Faraday cage. The magnetic covert channels such as ODINI and MAGNETO can be used to leak the private keys from an air-gapped wallet in a matter of minutes.  

\subsection{Optical}
The private key can be exfiltrated from air-gapped wallets via optical signals. The signals can be received by a nearby cameras, e.g., a webcam, smartphone, or security camera with a line-of-sight with the air-gapped computer. Few optical covert channels which are relevant to our attack model have been proposed over the years.

\subsubsection{Keyboard LEDs}
Loughry introduced the use of PC keyboard LEDs (caps-lock, num-lock, and scroll-lock) to exfiltrate binary data in an optical way \cite{loughry2002information}. The main drawback of this method is that it is not fully covert. Since keyboard LEDs don't usually blink the user can easily detect the transmission. 

\subsubsection{Hard-disk-drive LEDs}
In 2017, Guri et al presented LED-it-GO, a covert channel that uses the hard drive (HDD) indicator LED in order to exfiltrate data from air-gapped computers \cite{Guri2017}. The same authors presented a method for data exfiltration from air-gapped networks via router and switch LEDs \cite{guri2017xled}. In the case of HDDs and routers, the devices blink frequently; hence, transmissions performed via these channels will not raise the user's suspicious. The router LEDs are less relevant in the case of air-gapped wallets, unless the air-gapped wallets are maintained in an internal network with switches or routers.

The optical covert channels described above can be used to leak the private keys from an air-gapped wallet to nearby cameras in a few seconds. 

\subsubsection{Invisible image (VisiSploit) / QR stenography}
In some air-gapped wallets (e.g., BitKey \cite{BitKeySe86:online}) the signed transaction can be scanned from the screen rather than copied to removable media. The signed transaction is shown in a form of QR code on the computer display and can be scanned with a standard smartphone. Guri et al showed that data can be leaked optically through fast blinking images or low contrast invisible QR code projected on the LCD screen \cite{guri2016optical}. The QR code is invisible to humans but can be reconstructed by a snapshot taken by the smartphone camera. In our case, the private keys are covertly projected on the screen along with the QR code of the signed transaction. When the user scans the visible QR code, the invisible private keys are also scanned. 

\begin{figure}[t]
	\centering
	\includegraphics[width=\linewidth]{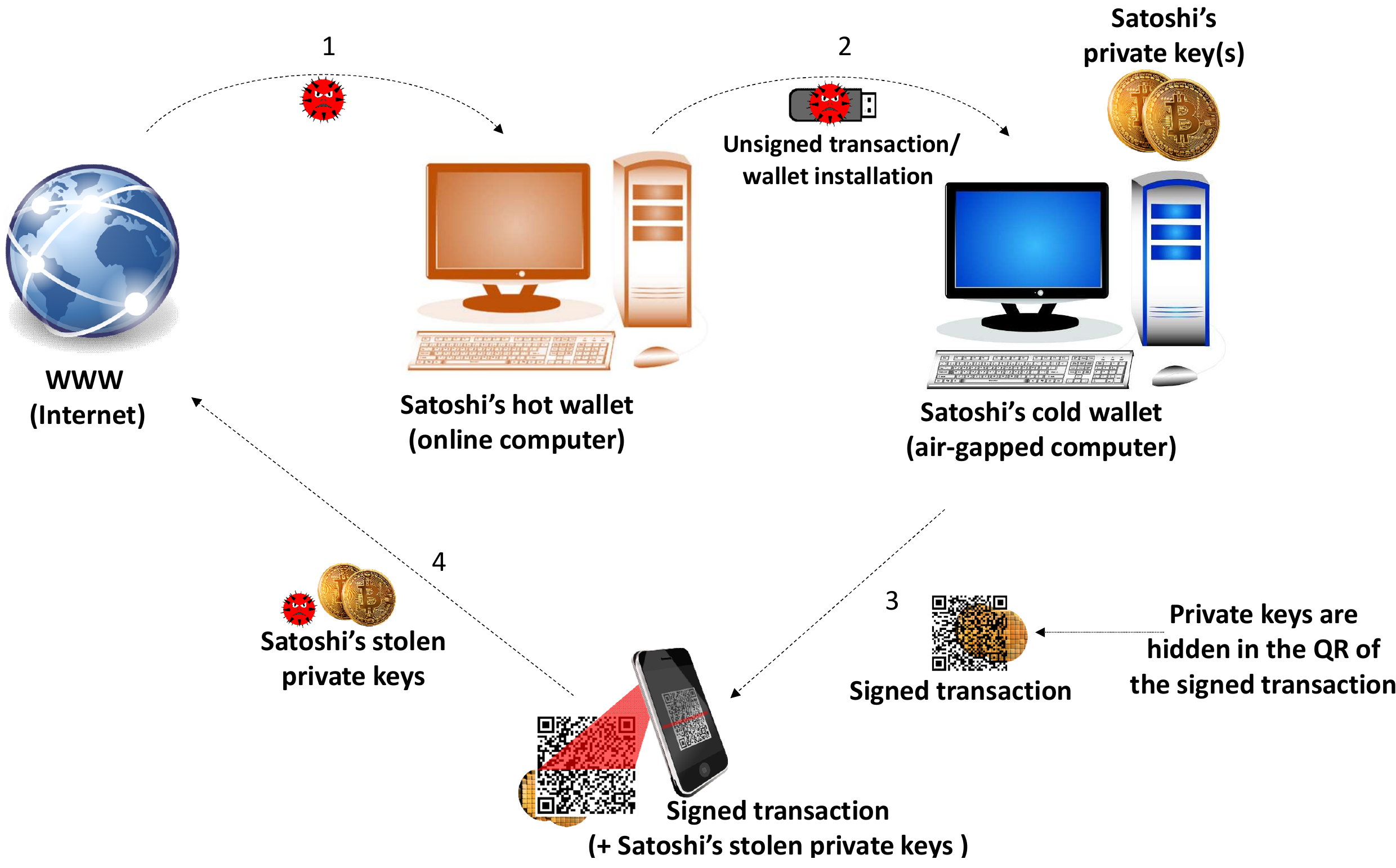}
	\caption{\textbf{Exfiltration of an the private keys during the transaction signing process. Satoshi's private keys are hidden in the signed transaction QR code. When Satoshi scans the QR code, the private keys are extracted and sent to the attacker through the Internet}}
	\label{fig:4}
\end{figure}

Another option is to hide the private key data within QR codes to establish a \textit{stenography} based covert channel \cite{cucurull2014qr}. Using this method, the private key (or part of it) is covertly embedded within the legitimate QR code of the signed transaction. After the signed transaction is scanned by the smartphone, the private keys are extracted and sent to the attacker. This covert channel is illustrated in Figure \ref{fig:4}.

\subsection{Acoustic}
In acoustic covert channels the private keys are exfiltrated via inaudible sound waves. Hanspach \cite{hanspach2014covert} show how to maintain an ultrasonic covert channel between air-gapped laptops equipped with speakers and microphones. He established communication between two computers located ~19 meters apart and achieved a bit rate of 20 bit/sec. Using the same method, Deshotels \cite{deshotels2014inaudible} showed that data can be transferred from computer to smartphone via ultrasonic waves. All of the aforementioned ultrasonic attacks are relevant to environments in which the computers are equipped with both speakers and microphones. The ultrasonic communication can be used to leak the  private keys from an air-gapped wallet to nearby PC or smartphone in a few seconds. 

\subsubsection{Ultrasonic (speaker-to-speaker communication)}
In many IT environments desktop computers are not equipped with microphones.
To overcome this limitation, Guri et al presented MOSQUITO \cite{guri2018mosquito} a malware that covertly turns headphones, earphones, or simple earbuds connected to a PC into a pair of  microphones, even when a standard microphone is not present. Using this technique they established so-called speaker-to-speaker ultrasonic communication between two or more computers in the same room. 
This attack is useful when the air-gapped computer is located in the same room with a microphone-less Internet-connected computer that is equipped with passive loudspeakers or headphones. It can be used to leak the private keys in a few seconds.
 
\subsubsection{Fansmitter: Computer fan noise}
In 2016, Guri et al introduced Fansmitter, a malware which facilitates the exfiltration of data from an air-gapped computer via noise generated from the computer fans \cite{guri2016fansmitter}. In this method, the air-gapped computer does not need to be equipped with loudspeakers, and the data could be leaked through acoustic signals generated from the computer fan.  

\subsubsection{Diskfiltration: hard-disk-drive noise}
Guri et al also presented a method dubbed DiskFiltration that uses the acoustic signals emitted from the hard disk drive (HDD) to exfiltrate data from air-gapped computers \cite{guri2017acoustic}. Similar to the previous attack, the air-gapped computer does not need to be equipped with loudspeakers, and the data could be leaked through acoustic noise generated by the HDD.  

The Fansmitter and Diskfiltration methods can be used to leak the private keys from an air-gapped wallet in a few minutes.

\begin{table*}[!t]
	\centering
	\caption{The air-gap covert channels relevant for private keys exfiltration}
	\label{tableall}
	\begin{tabular}{@{}llll@{}}
		\toprule
		\textbf{Type}                                  & \textbf{Method}                                                                                                                                                               & \textbf{Recevier}                                                                                        & \textbf{256-bit key }                             \\ \midrule
		\multicolumn{1}{|l|}{Physical}        & \multicolumn{1}{l|}{Removable and external media (E.g., USB flash drives)}                                                                                                        & \multicolumn{1}{l|}{Computer}                                                                   & \multicolumn{1}{l|}{\textless0.01 sec}   \\ \midrule
		\multicolumn{1}{|l|}{Electromagentic} & \multicolumn{1}{l|}{AirHopper (FM signals emitted from the video cable \cite{guri2017bridging,guri2014airhopper})}                                                                                             & \multicolumn{1}{l|}{Mobile phone}                                                               & \multicolumn{1}{l|}{\textless1 sec}      \\ \midrule
		\multicolumn{1}{|l|}{Electromagnetic} & \multicolumn{1}{l|}{GSMem (cellular interferences emitted from the CPU-RAM bus) \cite{guri2015gsmem}}                                                                                     & \multicolumn{1}{l|}{Mobile phone}                                                               & \multicolumn{1}{l|}{$\sim$300 sec}       \\ \midrule
		\multicolumn{1}{|l|}{Electromagnetic} & \multicolumn{1}{l|}{RADIoT (radio signals generated by embedded and IoT devices) \cite{RADIoTguri}}                                                                                    & \multicolumn{1}{l|}{\begin{tabular}[c]{@{}l@{}}Mobile phone/\\ Dedicated receiver\end{tabular}} & \multicolumn{1}{l|}{$\sim$1-50 sec}      \\ \midrule
		\multicolumn{1}{|l|}{Electric}        & \multicolumn{1}{l|}{PowerHammer (data exfiltrated thorough the power lines) \cite{2018powerhammer}}                                                                                         & \multicolumn{1}{l|}{Dedicated receiver}                                                         & \multicolumn{1}{l|}{$\sim$ 30-300 sec}   \\ \midrule
		\multicolumn{1}{|l|}{Magnetic}        & \multicolumn{1}{l|}{\begin{tabular}[c]{@{}l@{}}MAGNETO (magnetic signals generated by the CPU to smartphone) \cite{guri2018magneto} \\ ODINI (magnetic signals generated by the CPU) \cite{guri2018odini}\\ HDD (Magnetic signals emitted from the HDD) - laptops \cite{matyunin2016covert}\end{tabular}} & \multicolumn{1}{l|}{Mobile phone}                                                               & \multicolumn{1}{l|}{$\sim$70-1000 sec}   \\ \midrule
		\multicolumn{1}{|l|}{Acoustic}        & \multicolumn{1}{l|}{Ultrasonic (generated by loudspeakers) \cite{hanspach2014covert}}                                                                                                          & \multicolumn{1}{l|}{\begin{tabular}[c]{@{}l@{}}Computer/ \\ Mobile phone\end{tabular}}          & \multicolumn{1}{l|}{$\sim$1-20 sec}      \\ \midrule
		\multicolumn{1}{|l|}{Acoustic}        & \multicolumn{1}{l|}{MOSQUITO (speaker-to-speaker ultrasonic communication) \cite{guri2018mosquito}}                                                                                          & \multicolumn{1}{l|}{Computer}                                                                   & \multicolumn{1}{l|}{$\sim$2-20 sec}      \\ \midrule
		\multicolumn{1}{|l|}{Acoustic}        & \multicolumn{1}{l|}{Fansmitter (acoustic signals generated by the CPU/chassis fans) \cite{guri2016fansmitter}}                                                                                 & \multicolumn{1}{l|}{\begin{tabular}[c]{@{}l@{}}Computer/\\ Mobile phone\end{tabular}}           & \multicolumn{1}{l|}{$\sim$1000-2000 sec} \\ \midrule
		\multicolumn{1}{|l|}{Acoustic}        & \multicolumn{1}{l|}{Diskfiltraition (acoustic signals generated by the HDD actuator arm) \cite{guri2017acoustic}}                                                                            & \multicolumn{1}{l|}{\begin{tabular}[c]{@{}l@{}}Computer/ \\ Mobile phone\end{tabular}}          & \multicolumn{1}{l|}{$\sim$100-200 sec}   \\ \midrule
		\multicolumn{1}{|l|}{Optical}         & \multicolumn{1}{l|}{Keyboard LEDs \cite{loughry2002information}}                                                                                                                                   & \multicolumn{1}{l|}{\begin{tabular}[c]{@{}l@{}}Local camera \\ (e.g., webcam)\end{tabular}}     & \multicolumn{1}{l|}{$\sim$50-100 sec}    \\ \midrule
		\multicolumn{1}{|l|}{Optical}         & \multicolumn{1}{l|}{Hard disk drive LEDs (LED-it-GO) (optical signals by HDD indicator LED) \cite{Guri2017}}                                                                         & \multicolumn{1}{l|}{\begin{tabular}[c]{@{}l@{}}Local camera\\ (e.g., webcam)\end{tabular}}      & \multicolumn{1}{l|}{$\sim$10-100 sec}    \\ \midrule
		\multicolumn{1}{|l|}{Optical}         & \multicolumn{1}{l|}{Invisible images on screen \cite{guri2016optical}}                                                                                                         & \multicolumn{1}{l|}{Mobile phone}                                                               & \multicolumn{1}{l|}{A snapshot}          \\ \midrule
		\multicolumn{1}{|l|}{Optical}         & \multicolumn{1}{l|}{QR code steganography \cite{cucurull2014qr}}                                                                                                                           & \multicolumn{1}{l|}{Mobile phone}                                                               & \multicolumn{1}{l|}{A snapshot}          \\ \bottomrule
	\end{tabular}
\end{table*}

\subsection{Thermal}
In 2015, Guri et al presented  BitWhisper \cite{guri2015bitwhisper}, a thermal covert channel allowing an attacker to establish bidirectional communication between two adjacent air-gapped computers via temperature changes. The heat is generated by the CPU/GPU of a standard computer and received by temperature sensors that are integrated into the motherboard of the nearby computer. Due to the low bit rate we consider this method as a less relevant alternative for private key exfiltration.

\subsection{Other Techniques}
There are other air-gap covert channels that have been suggested over the years which requires a  hardware receivers or transmitters as a part of the attack. We consider these covert channels as less feasible for the attack model described in this paper. For example, in 2016, Guri et al presented USBee, a malware that uses the USB data buses to generate electromagnetic signals from a desktop computer \cite{guri2016usbee}. Similarly, researchers also proposed using GPIO ports of printers to generate covert radio signals for the purpose of data exfiltration \cite{funtenna86:online}. Both attacks require a dedicated RF receiver in the area. Lopes  presented a covert channel based on a malicious hardware component with implanted IR LEDs \cite{lopes2017platform}. However, in
this method the attacker must find a way to attach the
compromised hardware to the target computer. In 2017, Guri et al presented aIR-Jumper, a malware that uses security cameras and their IR LEDs to covertly communicate with air-gapped networks from a distance of hundreds of meters \cite{guri2017air}. This method is relevant only to corporate networks where surveillance cameras are installed.

Table \ref{tableall}. summarizes the relevant air-gap covert channels along with the estimation of time it takes to leak a 256-bit private key in each covert channel.



\section{Countermeasures}
\label{sec:countermeasures}
Many of the countermeasures for air-gap covert channels are adapted from standards and regulations for governmental and military organizations. Although some of the regulations are restrictive for personal users, they can be employed to some extent for the maintenance of air-gapped wallets. 
\subsection{Infiltration}
Anti-virus programs (AVs), host-based intrusion detection systems (HIDs) and host-based intrusion prevention systems (HIPs) may be used to prevent the initial infection of the air-gapped wallet with malicious code. Modern AVs may employ static scanning and runtime analysis to every file stored on the removable media device. However, malware authors have repeatedly proven that they can successfully bypass AVs, HIDs and HIPs by using zero-day vulnerabilities and employing stealth and evasion techniques \cite{25ofneww53:online},\cite{NewCrypt8:online},\cite{W32Dapro94:online},\cite{Sporathe68:online},\cite{CiscosTa82:online},\cite{ShadowPa49:online},\cite{Bewareof42:online},\cite{Kaspersk99:online},\cite{AFannyEq7:online},\cite{langner2011stuxnet}.
 
\subsection{Exfiltration}
It is possible to detect and block some covert channels presented in this paper using behavioral analysis. For example, hooking system resources and tracing the use of suspicious APIs \cite{guri2015gsmem,guri2015bitwhisper,guri2018mosquito} have been suggested for identifying intentional electromagnetic, acoustic, thermal, or optical transmissions. In this approach behavioral analysis, machine learning, and anomaly detection techniques may be used to detect the presence of covert channels and raise alerts. As noted in previous work on this topic, such forms of behavioral detection inherently suffer from high false positive rates  \cite{guri2015gsmem,guri2015bitwhisper,guri2018mosquito}.
\subsubsection{Policy-based countermeasures}
At the policy level it is possible to define a radius around the air-gapped wallet in which computers, smartphones, cameras, and other receivers are not allowed to cross. This approach is also known as red/black isolation, and refers to a physical separation between systems that may carry information with different levels of classification \cite{Guri:2018:BAM:3200906.3177230}. However, such measures might not be practical for private users. In addition, some air-gapped wallets  intentionally utilize smartphones for the transfer of transactions between cold and hot wallets \cite{BitKeySe86:online}.     
\subsubsection{Hardware-based countermeasures}
A basic hardware-based countermeasure scheme involves shielding computers with metallic materials to prevent electromagnetic radiation from leaking from the shielded equipment. Shielding can limit the effective range of many electromagnetic-based attacks. However, it is less suitable for private users due to the maintenance required and cost. When a highly valuable wallet is involved, a signal jamming approach might be taken. In this approach, a specialized hardware transmitter continuously generates random noises that interfere with potential transmissions from the wallet. Jamming is primarily used to block of electromagentic and acoustic signals \cite{punal2012vanets}.

\section{Conclusion}
\label{sec:conclusion}
The threat of data exfiltration from air-gapped computers is often discussed in the context of sophisticated cyber-attacks. However, with the emergence of cryptocurrencies (e.g., bitcoin) and the accompanying need to secure private keys from online threats, it has been suggested that private users manage their cryptocurrency wallets offline in isolated, air-gapped computers. 

We show that despite the high degree of isolation of cold wallets, motivated attackers can steal the private keys out of the air-gapped wallets. With the private keys in hand, an attacker virtually owns all of the currency in the wallet. In the attack model presented, the attacker infiltrates the offline wallet, infecting it with malicious code. Then, by using air gap covert channels, attackers can jump the air-gap and leak the private keys to nearby online computers, smartphones, or cameras. We evaluate the exfiltration techniques, including physical, electromagnetic, electric, magnetic, acoustic, optical, and thermal. We present a chain of attack that allows an attacker to compromise an air-gapped wallet and exfiltrate the private keys from it. We demonstrate how bitcoin’s private keys are exfiltrated from an offline, air-gapped wallet in a matter of a few seconds, using electromagnetic and acoustic covert channels\footnote{https://cyber.bgu.ac.il/advanced-cyber/airgap}.

\bibliographystyle{ieeetran}
\bibliography{Wallet}

\end{document}